\documentclass[sigchi]{acmart}
\AtBeginDocument{%
  \providecommand\BibTeX{{%
    \normalfont B\kern-0.5em{\scshape i\kern-0.25em b}\kern-0.8em\TeX}}}

\copyrightyear{2023} 
\acmYear{2023} 
\setcopyright{rightsretained} 
\acmConference[CHI '23]{Proceedings of the 2023 CHI Conference on Human Factors in Computing Systems}{April 23--28, 2023}{Hamburg, Germany}
\acmBooktitle{Proceedings of the 2023 CHI Conference on Human Factors in Computing Systems (CHI '23), April 23--28, 2023, Hamburg, Germany}\acmDOI{10.1145/3544548.3581075}
\acmISBN{978-1-4503-9421-5/23/04}

\newcommand{\multilinecell}[2][t]{%
  \begin{tabular}[#1]{@{}l@{}}#2\end{tabular}}

\begin{document}

\title[Ignore, Trust, or Negotiate]{Ignore, Trust, or Negotiate: Understanding Clinician Acceptance of AI-Based Treatment Recommendations in Health Care}

\author{Venkatesh Sivaraman}
\email{venkats@cmu.edu}
\orcid{0000-0002-6965-3961}

\affiliation{%
  \institution{Carnegie Mellon University}
  \streetaddress{5000 Forbes Ave}
  \city{Pittsburgh}
  \state{Pennsylvania}
  \country{USA}
  \postcode{15213}
}

\author{Leigh A. Bukowski}
\affiliation{%
  \institution{University of Pittsburgh, School of Medicine}
  \streetaddress{3550 Terrace Street}
  \city{Pittsburgh}
  \state{Pennsylvania}
  \country{USA}}

\author{Joel Levin}
\orcid{0000-0002-4095-4840}
\affiliation{%
  \institution{University of Pittsburgh}
  \streetaddress{3950 Roberto Clemente Dr}
  \city{Pittsburgh}
  \state{Pennsylvania}
  \country{USA}}
  \postcode{15260}
\email{joel.levin@pitt.edu}

\author{Jeremy M. Kahn}
\affiliation{%
  \institution{University of Pittsburgh and UPMC Health System}
  \streetaddress{3550 Terrace Street}
  \city{Pittsburgh}
  \state{Pennsylvania}
  \country{USA}}
\email{jeremykahn@pitt.edu}

\author{Adam Perer}
\orcid{0000-0002-8369-3847}
\affiliation{%
  \institution{Carnegie Mellon University}
  \streetaddress{5000 Forbes Ave}
  \city{Pittsburgh}
  \state{Pennsylvania}
  \country{USA}
  \postcode{15213}
}
\email{adamperer@cmu.edu}

\renewcommand{\shortauthors}{Sivaraman et al.}

\begin{abstract}
Artificial intelligence (AI) in healthcare has the potential to improve patient outcomes, but clinician acceptance remains a critical barrier. We developed a novel decision support interface that provides interpretable treatment recommendations for sepsis, a life-threatening condition in which decisional uncertainty is common, treatment practices vary widely, and poor outcomes can occur even with optimal decisions. This system formed the basis of a mixed-methods study in which 24 intensive care clinicians made AI-assisted decisions on real patient cases. We found that explanations generally increased confidence in the AI, but concordance with specific recommendations varied beyond the binary acceptance or rejection described in prior work. Although clinicians sometimes ignored or trusted the AI, they also often prioritized aspects of the recommendations to follow, reject, or delay in a process we term ``negotiation.'' These results reveal novel barriers to adoption of treatment-focused AI tools and suggest ways to better support differing clinician perspectives.
\end{abstract}




%   \includegraphics[width=\textwidth]{figures/visualization_conditions.png}
%   \caption{Clinicians made decisions on four patients, assisted by an AI treatment recommendation in three cases: (a) }

\maketitle

\section{Introduction}

Artificial intelligence (AI) in health care promises to improve outcomes, reduce costs, and save clinicians time and effort. Yet at present, even the most encouraging AI solutions face significant obstacles to deployment and acceptance in real-world clinical settings \cite{Panch2019}. AI-based tools that seek to improve decision-making across diverse deployment contexts must produce recommendations that are both acceptable to health care providers and transparent in the case of errors \cite{Futoma2020,Vasey2022}. In addition, health care providers generally consider themselves to be content experts in their fields, and they are naturally skeptical of decision aids that may limit their autonomy and challenge their sense of identity \cite{shah2019ai}. These issues have motivated studies that aim to evaluate and improve the human-AI collaborative system in health care \cite{Henry2022,Jacobs2021psych,Lee2021}, often drawing on insights from interpretable AI \cite{Schoonderwoerd2021,Wang2020,Lai2021}. By helping clinical experts understand the conditions in which AI predictions fail, clinical decision support (CDS) systems could help produce AI-assisted decisions that are better than those made by humans or algorithms alone, improving patient outcomes.

%% Other reasons why generalizable accuracy is important for adoption:
%% - (People) won't adopt algorithmic tools that they observe making significant errors \cite{Dietvorst2015}

Despite these efforts, effective complementarity between humans and AI-based CDS has largely not yet been realized, in part because it is difficult for clinicians to calibrate their trust in newly developed AI systems. Experimental studies demonstrate that trust can be miscalibrated in both directions: experienced clinicians often dismiss AI recommendations regardless of quality, while novices over-rely on incorrect advice \cite{Gaube2021,Bayer2021}. Prior work has investigated several strategies to mitigate these discrepancies, including providing explanations of the process underlying a given recommendation \cite{Alam2021,Bussone2015,Wang2020}, communicating the uncertainty of predictions \cite{Tonekaboni2019,Zhang2020}, and familiarizing users with the AI's global strengths and weaknesses as identified from external validation \cite{Ghassemi2021,Cai2019}. However, none of these methods appear to work universally across contexts, particularly in health care; in some domains, they may inflate or undermine confidence \cite{Gaube2021,Jacobs2021psych,Zhang2020}, while in others they may be poorly-aligned with human decision-making processes \cite{Amann2022,Ghassemi2021}. In order to cultivate an appropriate level of reliance, interpretable AI systems must account for decision-maker characteristics, task complexity, AI performance, and other factors in ways that are not yet fully understood \cite{Markus2021,Schoonderwoerd2021,Lai2021}.

Importantly, most work in this area has focused on \textit{diagnostic} systems—that is, systems designed to help clinicians make a clinical diagnosis (e.g., identifying cancer in a radiograph \cite{Calisto2021}) or predict a future clinical event (e.g., clinical deterioration \cite{Henry2022,Sendak2020facct}). These systems have the benefit of a known or expert-annotated ``ground truth'' upon which algorithms can be developed and calibrated. As such, they aim to improve clinical care by reducing diagnostic errors, providing insight into the likelihood of future events, and minimizing cognitive burden.

A separate, emerging class of CDS systems is designed to deliver \textit{treatment} recommendations—for example, which type of chemotherapy to give to a cancer patient or whether or not to administer intravenous fluids to a hospitalized patient identified as being at-risk. In these systems, the ``best'' decisions are often difficult to identify, either due to a lack of clinical evidence \cite{ebell2017good,wilson2011appraisal} or due to expert disagreement on the best course of action \cite{eddy1990clinical,rubenfeld2001understanding}. Therefore, treatment-focused AI models often seek to discover optimal decisions by correlating treatments with their average effects on patient outcomes, a task which is significantly more challenging than diagnosis or prediction but which has the potential to more powerfully impact patients. 

While clearly different from a modeling perspective, a key question is whether AI tools that do not have a clear correct decision require different approaches for \textit{human-AI interaction design}. For diagnostic aids and risk-assessment tools, AI-assisted decision-making behavior is often conceptualized as taking place at a single point in time (e.g. a physician encounter) and as involving a limited number of options (e.g. either agreement or disagreement with a diagnosis) \cite{Wang2021brilliantAI,Henry2022,Tschandl2020,Jussupow2020,Cai2019,Calisto2021}. In contrast, decisions about \textit{treatments} often span a wider range of options and take place across multiple time-points that can confound outcomes. For instance, a clinician may prescribe one treatment on the first visit, then observe that it is having little effect on the patient's condition and administer a different treatment upon the second visit. Determining the optimal decision in this context requires understanding both how past treatment decisions have affected the current patient state, and how future treatment decisions might influence the expected outcome. This complex reasoning task is well-known in causal modeling \cite{Parbhoo2022} and has been described in early-stage design studies in health care \cite{Kaltenhauser2020}. It has not, to our knowledge, been explored in the context of a functioning treatment decision support tool.

In this work, we sought to explore how clinicians interact with real AI-based treatment recommendations in a setting where sequential treatments can affect outcomes in complex ways \cite{Kaltenhauser2020}. Our clinical domain of interest was the intensive care unit (ICU), an environment characterized by acutely ill hospitalized patients and correspondingly dynamic, time-sensitive decisions. We designed and implemented an interactive CDS interface that delivers interpretable recommendations for treating sepsis, a life-threatening medical condition with relatively few existing evidence-based care protocols and substantial heterogeneity in treatment patterns among clinicians \cite{Cecconi2018}. The foundation of our CDS was an existing well-known AI model that could reduce patient mortality if followed \cite{Komorowski2018} but that has not been prospectively evaluated in a clinical setting. The resulting CDS system formed the basis for a think-aloud study with 24 clinicians, all of whom practice in the ICU and have experience treating sepsis. We aimed to understand their responses to the recommendations and explanatory evidence provided by the AI, including both how they perceived it to influence their decisions and how their actual treatment choices were affected.

A mixed-methods analysis of the think-aloud transcripts and structured decision responses showed that explanations improved clinicians' perceptions of the AI's usefulness and made them more confident in their own decisions, a finding that is consistent with prior literature \cite{Alam2021,Tschandl2020,Wang2020}. However, their overall rates of binary concordance with the AI recommendations did not appear to be affected by explanatory visualizations. Instead, analysis of participants' think-aloud decision processes revealed a more nuanced picture of individual decision-making than described in the literature thus far, involving four distinct behavior patterns with the AI:
\begin{enumerate}
    \item \textbf{Ignore,} in which the decision-maker is not affected by the AI recommendation in any decision;
    \item \textbf{Negotiate,} in which the decision-maker weighs and prioritizes individual aspects of the recommendation to follow or adjust;
    \item \textbf{Consider,} in which the decision-maker dichotomously defers to or overrides the recommendation; and
    \item \textbf{Rely,} in which the decision-maker accepts some part of the recommendation in every decision.
\end{enumerate}

\noindent These behavior patterns, particularly in the Negotiate group, indicate that recommendations for treatment decisions in the ICU may be subject to \textit{partial} forms of reliance that could impact the efficacy of chosen treatments in undetermined ways. Our results also pointed to ways in which the formulation of the model used for this study hindered clinicians from using it effectively, opening new directions for model improvement and evaluation. We discuss the implications of the behavior patterns and obstacles we observed on the further development of AI-based treatment decision support tools.

% [It is useful to draw a distinction between AI used to augment perception and diagnosis, and AI used for statistical risk prediction such as treatment effectiveness. (Insert prior distinction between these two approaches here...) AI models used for treatment recommendation are in some ways more akin to risk assessment tools, such as those being applied to criminal justice \cite{Johndrow2019} and child welfare \cite{Kawakami2022partnerships}, than to deep neural networks that can make binary diagnoses from medical images.]

% a chicken-and-egg validation problem: clinicians will not trust an untested system to make decisions, but the system cannot be validated unless clinicians use it regularly \cite{Yang2019}

\section{Background and Related Work}

\subsection{Sepsis Diagnosis and Treatment in the ICU}

Sepsis is a life-threatening medical condition that affects over 1.7 million adults in the United States each year and is the leading cause of death in hospitals \cite{CentersforDiseaseControlandPrevention2021}. Sepsis occurs when the body's response to an infection causes systemic inflammation and organ dysfunction \cite{Evans2021}, which can in turn lead to septic shock and death \cite{Shankar-Hari2016}. Sepsis is also the most costly condition in U.S. hospitals \cite{buchman2020sepsis}, and as such it represents a major target for quality improvement efforts at the local and national level \cite{hershey2017state}. Timely identification and appropriate management of sepsis is crucial to reducing mortality rates \cite{Evans2021}. Key diagnostic strategies include frequent clinical assessments, blood cultures to identify pathogens, and measurement of laboratory values that may indicate infection; treatment strategies include control of the infectious source with antibiotics or antivirals, intravenous (IV) fluids to maintain appropriate fluid balance, and vasopressors (such as norepinephrine) to maintain appropriate blood pressure \cite{Evans2021}.

Sepsis has long been a focus of AI research; however, nearly all of this research is devoted to early identification and diagnosis. Multiple machine learning algorithms exist for mining hospital electronic health record data to identify patients with sepsis \cite{teng2020review,moor2021early,sadasivuni2022fusion}. These systems are generally accurate, and several hospital systems have already implemented algorithmic early warning systems for sepsis \cite{Sendak2020,Henry2022}. However, the implementation of these systems has not tended to affect treatment decisions or patient outcomes \cite{Hooper2015}. Conceptually, early warning systems may fall short of their goal of improving the quality of care when they fail to provide information that is both novel and actionable. 

In contrast, relatively little AI research has focused on sepsis treatment. Guidelines for treating sepsis in the ICU are continually evolving \cite{Evans2021}, and although individual treatment decisions at specific time points (e.g., whether to give fluids or vasopressors) are certainly highly influential on patient outcomes \textit{on average}, the physiological complexity of sepsis renders the influence of treatments on individual outcomes largely unknowable. As such, recommendations face significant challenges in translation to wider clinical practice \cite{Stoneking2011}, resulting in substantial variability in care practices \cite{barbash2019national} and continued high mortality levels \cite{stevenson2014two}.

Machine learning approaches for sepsis treatment aim to standardize and improve sepsis care by leveraging historical patient trajectories. The most prominent example of this approach is the AI Clinician developed by Komorowski et al. \cite{Komorowski2018}, and it is the model that forms the basis for our clinician-facing work. By improving the consistency and timeliness around treatment with IV fluids and vasopressors, sepsis treatment models such as the AI Clinician have immense potential to reduce mortality (from around 13\% to around 5\%, according to \cite{Futoma2018}). However, for this potential to be realized, clinicians actually have to act on the AI recommendations at the bedside. Indeed, studies evaluating the effects of these predictions have only considered retrospective data and not how (or if) such tools might be utilized by human clinicians. Because model recommendations are impactful only if they are implemented, it is critical to understand how a model like the AI Clinician might be integrated into an ICU clinician's workflow, and whether human-AI collaboration can indeed outperform unaided human clinicians.

\subsection{Explainability, Interpretability and Decision-Making}

The design of explainable and interpretable ML-based tools has become a major focus of research in the HCI community \cite{Suresh2021,Ehsan2021,Zhang2020}. While early efforts in explainable AI (XAI) focused on feature-based explanations, current conceptions of interpretability comprise a wider range of techniques, including uncertainty and confidence metrics \cite{Schaekermann2020}, nearest-neighbors \cite{Islam2021}, and counterfactuals \cite{Zytek2021}. In concert with human-centered design methods, these technical approaches can be integrated into algorithmic systems with the intent of improving trust and human-AI team performance \cite{Hong2020,Schoonderwoerd2021,Bansal2020}.

However, there remain substantial challenges in designing and validating interpretable AI systems, particularly in high-stakes decision-making domains such as health care. Model explanations themselves can be prone to issues such as over-sensitivity to input values or giving seemingly-sensible explanations for incorrect predictions \cite{Slack2021}. When explanations are presented to decision-makers alongside predictions, a line of studies ranging from Bussone et al. \cite{Bussone2015} to more recent work \cite{Wang2020,Zhang2020,Chromik2021} has shown that these explanations tend to increase trust in the model even when it is unwarranted. Explanations can also interact with reasoning fallacies such as confirmation and availability bias \cite{Jussupow2020,Dobler2019}, but mitigating these effects requires knowledge of a normatively correct reasoning process \cite{Wang2019} that may not always be available.

In the translation of promising AI tools into a real-world setting, the evaluation of decision quality poses its own challenges. While expert consensus can serve as a useful proxy for the ground truth \cite{Jacobs2021psych}, the accuracy of a real-world decision is often fundamentally unknowable and contentious \cite{Kawakami2022partnerships}. Taking an alternative strategy, some AI systems instead strive to provide clinicians with useful non-prescriptive information, such as highlighting informative parts of a medical image \cite{Esteva2021}, displaying information from similar historical cases \cite{Cai2019imageretrieval}, or identifying patients at high risk of future deterioriation \cite{Henry2022,Sendak2020facct}. These approaches serve to focus attention without making specific recommendations, thereby indirectly improving decisions but also making the AI algorithms more ignorable (potentially reducing benefit).

The present study was specifically designed to address the challenges described above: imperfect and biased models, potentially misleading and hard-to-interpret explanations, and in particular the lack of an objective ground truth. Rather than evaluating decision-making along a single axis of quality, we used a combination of behavioral and attitudinal measures \cite{Scharowski2022} to understand how an imperfect AI system would affect its users within a high-stakes environment in which the correct decision is unknowable in real-time.

\subsection{Clinician Perceptions of Decision Support Tools}

While ML-based tools for clinical decision-making have great potential utility, they face the combined challenges of building effective XAI as well as broader obstacles to adoption of CDS tools. Yang et al. \cite{Yang2019} describe difficulties in gaining acceptance from expert clinicians without formal validation of the tool, as well as the inherent challenge of situating CDS at the right time and place for decision-making. Similarly, Cai et al. \cite{Cai2019} emphasize clinicians' need to understand the overall design and validation of the CDS before they can trust it on individual instances. Studies of deployed AI systems by Beede et al. \cite{Beede2020} and Wang et al. \cite{Wang2021brilliantAI} have identified clinician frustrations with the added workload of using a CDS, particularly when those systems do not adequately complement their expertise. Early-stage studies of CDS tools have also found that acceptance of AI recommendations is often more strongly determined by the clinician's expertise than by the quality of the recommendation \cite{Tschandl2020,Gaube2021}.

On the other hand, a few deployed systems have met with success and clinician acceptance. For example, AI-driven CDS systems for image-based diagnosis have been increasingly accepted as tools to reduce clinician burden and prioritize attention \cite{Tschandl2020,Beede2020}. Related to sepsis, early-warning systems such as Sepsis Watch and the Targeted Real-time Early Warning System (TREWS) have been accepted by clinicians at the hospitals where they are deployed \cite{Sendak2020,Henry2022}, despite being initially met with ambivalence \cite{Ginestra2019}. These tools may have been readily accepted because (1) they could be rigorously validated using ground-truth data, (2) they ultimately helped coordinate providers and prioritize care \cite{Henry2022} rather than directly replacing clinical judgment; and (3) acceptance does not rely on clinician behavior change and they can therefore be easily ignored by untrusting clinicians. However, while these diagnosis-focused tools are a promising model for AI-based CDS, they represent only one of many points in the care workflow in which complex decisions may be needed. 

In particular, relatively few studies have examined the acceptability of AI-generated \textit{treatment recommendations}: Jacobs et al. used a mock AI system to evaluate clinician decisions on antidepressant selection \cite{Jacobs2021psych}, while Yang et al. presented clinicians with projected survival curves conditioned on a device implantation decision \cite{Yang2019}. Kaltenhauser et al. examined intravenous fluid administration in intensive care, although their study was more focused on understanding decision-making without an AI rather than the influence of AI on treatment decisions \cite{Kaltenhauser2020}. In contrast to diagnoses, which are relatively straightforward to learn from historical data, treatment recommendations in CDS have predominantly been derived from clinical best-practice guidelines rather than AI \cite{Jones2019,Beauchemin2019}. However, broad best-practice guidelines are of limited utility at the bedside because their recommendations fail to account for patient-level variation and the interaction between multiple variables over time \cite{klompas2020current}. Machine learning approaches have the potential to deliver treatment recommendations that are more specific and personalized, but how they will be received by clinicians remains an open question.

\section{Design of an Interactive AI-Driven CDS System}

As an initial step towards bringing AI-based recommendations to clinical practice for sepsis treatment, we designed and implemented an interactive patient trajectory visualization tool called the AI Clinician Explorer. This tool serves as both an exploratory tool for historical patient data and as an interface for a real treatment recommendation model developed from retrospective clinical data (the AI Clinician). The following sections describe the underlying model as well as the design of the front-end visualization system.

\subsection{Reinforcement Learning for Sepsis Treatment}\label{sec:model-description}
\begin{figure}
    \centering
    \includegraphics[width=0.48\textwidth]{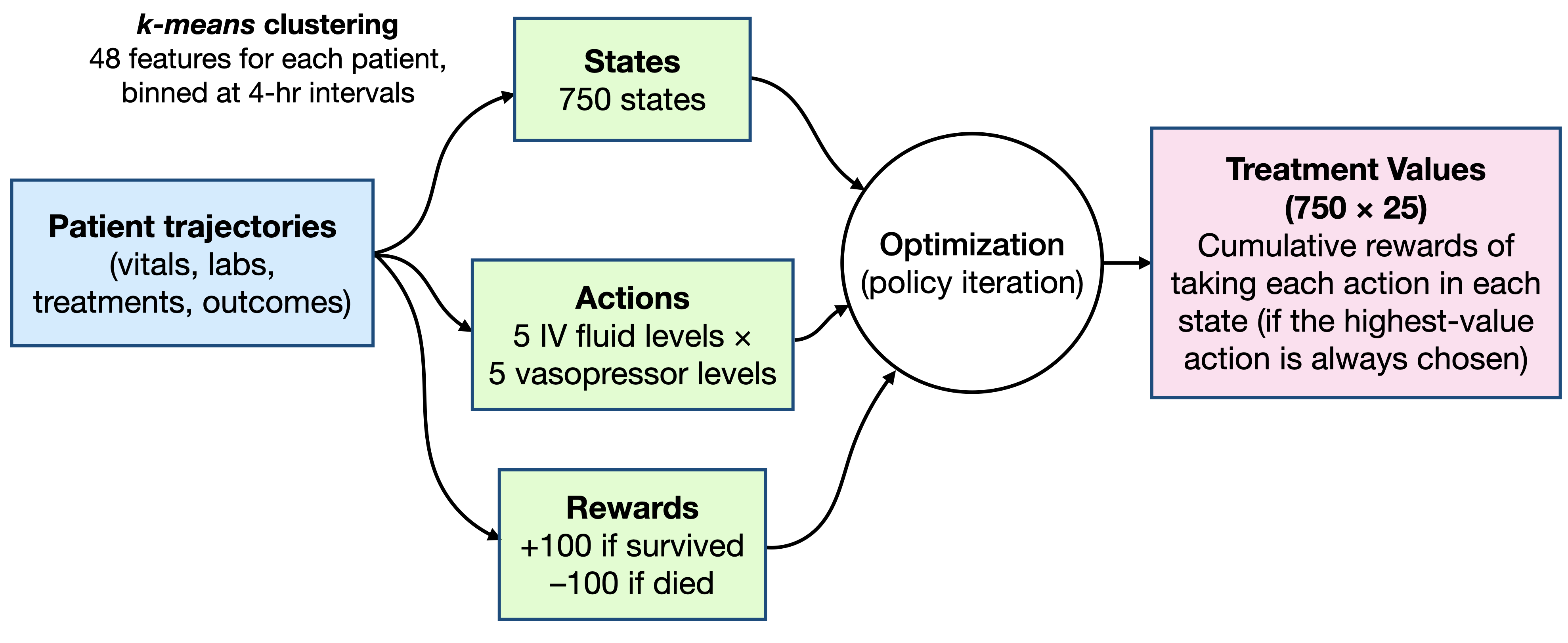}
    \caption{Overview of the AI Clinician's training methodology, summarized from \cite{Komorowski2018}. The model takes as input a set of historical trajectories comprising patient vitals, labs, and treatments discretized at 4-hour intervals. Each timestep is represented as one of 750 different states (determined using clustering) followed by one of 25 possible treatment actions. The output of the model is a set of treatment values (or Q-values), which estimate the future rewards that would be obtained from taking a given action. The \textit{policy} that the AI Clinician would follow is to take the action with the largest value estimate in each state.}
    \label{fig:ai_clinician_training}
\end{figure}

Unlike many AI problem formulations which treat the patient as a static data point on which to make a prediction, sepsis management in the ICU requires a dynamic approach that considers the changing state of the patient over time. In particular, models need to account for the fact that long-term outcomes, such as mortality, may not be the direct result of a single action but rather a series of actions over an evolving trajectory. The AI Clinician \cite{Komorowski2018} addresses these challenges by applying a reinforcement learning (RL) strategy. Like most RL approaches in health care, the AI Clinician works by modeling a patient trajectory as a sequence of memoryless states derived from available biometric signals  (vital signs, lab values, etc.). At each timestep, the agent—either a clinician or a model—can choose from a predefined set of actions, which then results in the agent receiving a numerical reward (or penalty). As shown in Fig. \ref{fig:ai_clinician_training}, the AI Clinician uses $k$-means clustering to define 750 possible patient states, then applies an algorithm called policy iteration to determine which of 25 different treatment actions most optimally reduces mortality in each state. The model's actions represent 5 possible levels of IV fluids and 5 vasopressor dosages binned by quantiles, representing a substantial but non-exhaustive subset of the treatment choices that a human clinician might make.

The AI Clinician publication is widely known and highly influential in both the critical care and CDS communities \cite{Topol2019,Yu2023}. Yet it has also faced criticism because its recommendations often deviate from bedside clinicians' best understanding. This may be due to inherent biases in how patients' outcomes are weighted in the model's evaluation process, a problem known in RL as off-policy evaluation \cite{Jeter2019}. Additionally, more recently-developed techniques using deep neural networks may improve on the accuracy of the AI Clinician \cite{Peng2018,Yu2019,Futoma2018}. However, since these more recent methods also rely on off-policy evaluation, reliable benchmarks of their performance remain elusive. For this study, we chose Komorowski et al.'s approach because it is the best-known model of its kind, and therefore most likely to gain clinician acceptance in the absence of concrete evidence that any such AI model improves outcomes.

We replicated the AI Clinician's methodology using the publicly-available MIMIC-IV dataset \cite{johnson2020mimic} (a more recent version of the MIMIC-III dataset used by the original model developers), and provide the code on GitHub for future reproducibility\footnote{\url{https://github.com/cmudig/AI-Clinician-MIMICIV}}. MIMIC includes granular clinical data on all ICU admissions to a large academic medical center over a multi-year time period, and therefore is a unique resource for this project. The model was trained on a cohort of 18,143 patients who met standard diagnostic criteria for sepsis at some point during their ICU stay. We verified that the model performance on held-out data was similar to the original reported performance, as measured by a bootstrapped policy value estimate computed using weighted importance sampling\footnote{The accuracy of an RL policy cannot be computed directly on retrospective data because we cannot observe the outcomes of following the policy. Instead, weighted importance sampling (WIS) works by averaging the survival/mortality rewards associated with each trajectory, weighted by how similar the clinicians' actions were to the model predictions.}. Specifically, the model whose predictions were displayed had a policy value of $83.8$ (possible values range from $-100$ to $100$), while the values reported by Komorowski et al. on MIMIC-III ranged between 80 and 90 \cite{Komorowski2018}.

\subsection{Visualization System Design}

\begin{figure*}
    \centering
    \includegraphics[width=0.95\textwidth]{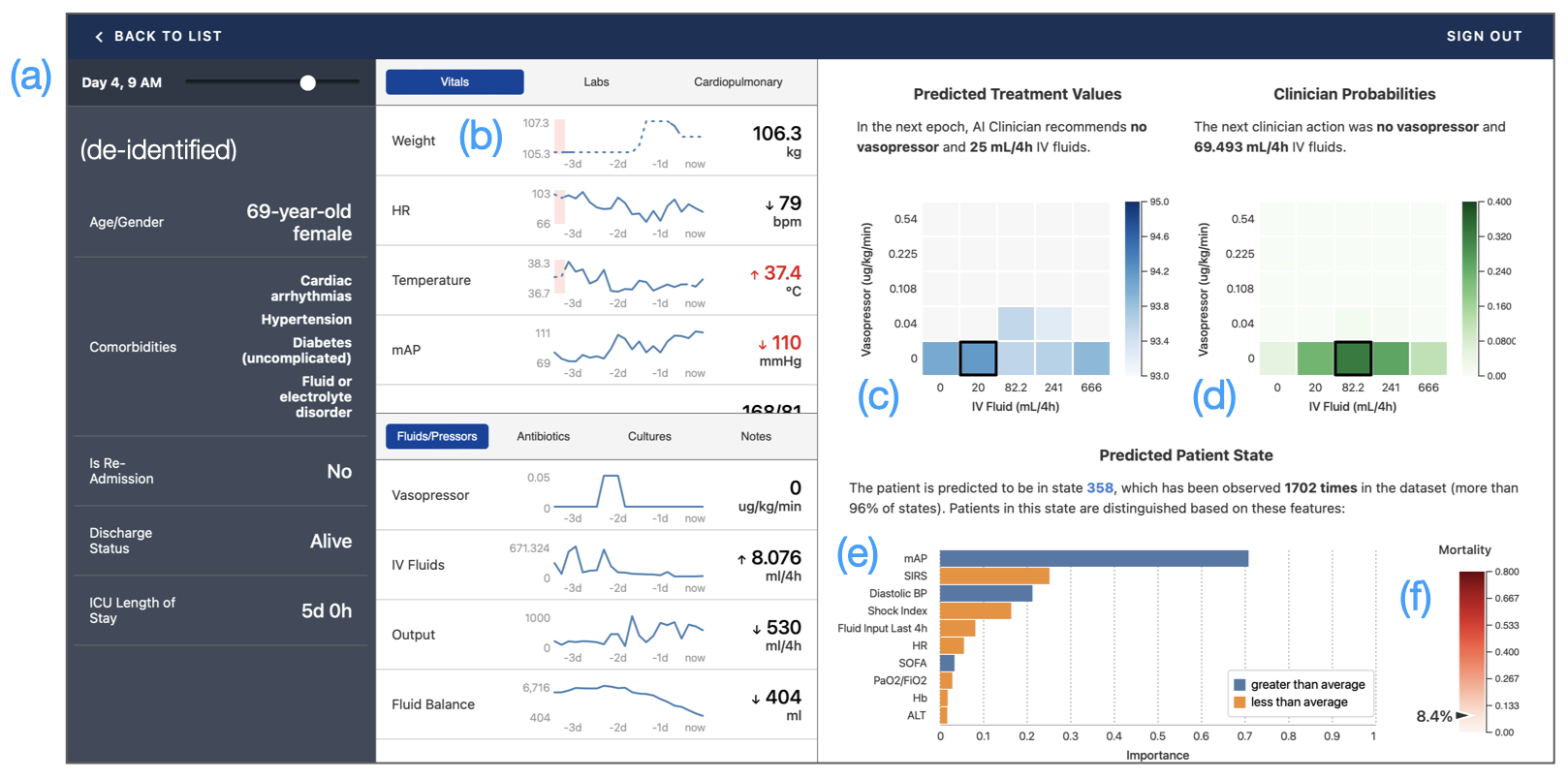}
    \caption{Main interface in the AI Clinician Explorer, designed to support browsing patient trajectories, interpreting model recommendations and comparing predicted treatment values against historical clinical actions. (a) The timestep control allows the user to step through the patient's ICU stay at 4-hour intervals. (b) The patient state is shown in small-multiple line charts, with abnormal values highlighted in red. (c) Heatmap showing the estimated value of taking each of the 25 possible actions ($Q$ values) from the current state. (Values are only estimated for actions with more than 5 observed clinician actions, as shown by the colors in the Clinician Probabilities plot.) (d) Probability of clinician actions for patients in the current state. (e) Description of the current patient state, as well as a chart showing most strongly contributing features according to SHAP. (f) Mortality rate of patients after being observed in this state.}
    \label{fig:ai_clinician_explorer}
\end{figure*}

We next developed a novel front-end visualization system which we term the AI Clinician Explorer. This system enables clinical experts to search for patients in the MIMIC-IV dataset, visualize their disease trajectories, and compare model predictions to actual treatment decisions delivered at the bedside. The AI Clinician Explorer was designed for use both as a tool for research and education on AI in sepsis, and as a starting point for an eventual clinician-facing interface for real-time decision-making in the live clinical environment. An initial prototype was created using inspiration from prior literature, notably ClinicalVis \cite{Ghassemi2018}. We then iterated on this design and tailored it for use by ICU clinicians, based on feedback from experienced ICU physicians and other experts in biomedical sciences, informatics, and psychology. The final system consists of the following primary components:

\textbf{Browse and filter patients.} 
The Patient Browser page helps users find cases of interest by allowing them to filter and sort a list of patient trajectories by a variety of task-specific metrics. The most straightforward of these include filters for age, gender, comorbidities, outcomes, and commonly-used disease severity scores (SOFA and SIRS). During the iterative development process, we identified a need to filter for specific actions and recommendations at a timestep level (i.e., the 4-hour time periods over which the model aggregates data and makes treatment recommendations), which is more granular than filtering at the patient-level. We added filter controls that allow the user to select from the 25 possible clinician and model actions on a pair of grids. This was used to identify timesteps in which, for example, clinicians tended to give IV fluids while the model recommended vasopressors.

\textbf{Visualize patient trajectory.}
The Patient Trajectory page, represented in Fig. \ref{fig:ai_clinician_explorer}, was designed to help clinicians quickly assess a patient's state throughout their ICU stay. Similar to ClinicalVis \cite{Ghassemi2018}, our trajectory visualizations communicate the patient's current vital signs and lab values numerically, and depict their trends over time using line charts. Abnormal values are highlighted in red, while trend arrows show changes in each value relative to the last 4 hours. Clinicians' feedback on the time-series charts indicated that the visualizations were highly usable, particularly compared to how data is currently presented in existing hospital-based electronic health records (EHRs). We also worked with the clinicians to reorder and regroup the features into semantic categories, which served to align the page's structure with standard reporting conventions and facilitate skimming.

\textbf{Compare model predictions and clinician actions.}
As described in Sec. \ref{sec:model-description}, the AI Clinician categorizes each patient to one of 750 states, and each state is associated with a predicted treatment recommendation. For each timestep in each patient trajectory, the interface displays heatmaps showing the AI Clinician's recommended action (Fig. \ref{fig:ai_clinician_explorer}c) and the distribution of historical clinician actions in that state (Fig. \ref{fig:ai_clinician_explorer}d). This pair of visualizations surfaced the insight that the AI Clinician often assigns similar treatment values to multiple actions rather than strongly preferring a single action. We therefore explored ways to present multiple treatment options during the study, resulting in the Alternative Treatments visualization condition.

\textbf{Interpret state clustering.}
Presenting explanations of the AI Clinician's predictions was a key aspect of both the patient browser interface and the clinician-facing study. A few explainability methods have been developed specifically for reinforcement learning (XRL) \cite{Madumal2020,Puiutta2020}; however, these methods generally either require different training methods (to create intrinsically interpretable policies) or accurate models of patient trajectory dynamics (to predict counterfactual outcomes). To align our work with both the existing AI Clinician model and prior XAI literature \cite{Wang2020,Zhang2020}, we opted to use standard XAI techniques to explain the state clustering, which has a major impact on the model output as it determines which groups of patients are recommended similar treatments. For each of the 750 possible states, we trained an XGBoost classifier \cite{chen2016xgboost} to predict whether a patient was in that state or not. We then used Shapley Additive Explanations (SHAP) \cite{Lundberg2017} to identify the features that most often contributed to patients being included in the state.\footnote{The choice of explanation technique can have a significant effect on what conclusions are drawn from feature importance charts. In our case, the combination of XGBoost and SHAP qualitatively yielded more parsimonious, clinically sensible explanations than other classifiers (random forests, SVMs) or explanation techniques (SVM coefficients, permutation importance).} These features are depicted in the State Interpretation chart, and were used in the Feature Explanation study condition.

We used the AI Clinician Explorer to select patients for our clinician-facing study, as described in greater detail in Sec. \ref{sec:case-selection}. The system also formed the basis of the interface that study participants used to make decisions. (In the study we controlled which model visualizations and time points participants saw rather than giving them access to the entire AI Clinician Explorer, thus better replicating the information set that would be available to clinicians in real life.) The tool is built using a Flask back-end, a Svelte front-end, and a database comprising BigQuery and Google Cloud Firestore components. The source code for the tool and study interface is available on GitHub to support future research\footnote{\url{https://github.com/cmudig/ai-clinician-explorer}}.

\section{Study Methods}\label{sec:methods}

We conducted a mixed-methods study to understand the challenges that clinicians face when attempting to incorporate AI advice, to explore how participants perceive their decision-making differently with AI support, and to evaluate the effect of explanations on acceptance. We explored the following research questions:

\begin{enumerate}
    \item How do AI-generated sepsis treatment recommendations and explanatory visualizations affect clinicians' perceptions of decision-making?
    \item How do visual explanations of model predictions affect acceptance of the AI's advice?
    \item What challenges do clinicians perceive in incorporating AI treatment recommendations into their decision-making?
\end{enumerate}

We recruited 24 practicing ICU clinicians from a large multi-hospital academic hospital system in the eastern United States. Our sample included three types of ICU clinicians representing the range of providers that make sepsis treatment decisions in the ICU: attending physicians, advanced practice providers (APPs), and critical care fellows in training. Attending physicians are the most senior clinicians in the ICU. APPs and fellows are generally less senior but still make independent decisions about sepsis treatments. Most participants were attending physicians, although their level of experience in the ICU varied significantly, as shown in Table \ref{tab:demographic_info}. Sessions were conducted on Zoom and lasted between 20 and 50 minutes; participants received 50 USD in compensation.

\begin{table}[]
    \centering
    \begin{tabular}{ccc}
    \toprule
    \textbf{Participant} & \textbf{Role} & \textbf{Years ICU Experience} \\
    \midrule
         P1 & APP & 1-2 \\
P2 & APP & 3-5 \\
P3 & Fellow & 3-5 \\
P4 & Fellow & 1-2 \\
P5 & Fellow & <1 \\
P6 & APP & >10 \\
P7 & Attending & 5-10 \\
P8 & Attending & 5-10 \\
P9 & Attending & >10 \\
P10 & Attending & 5-10 \\
P11 & Attending & >10 \\
P12 & Attending & >10 \\
         P13 & APP & >10 \\
P14 & Attending & >10 \\
P15 & Attending & 3-5 \\
P16 & Attending & >10 \\
P17 & Attending & >10 \\
P18 & Attending & 5-10 \\
P19 & APP & >10 \\
P20 & APP & 3-5 \\
P21 & Attending & 3-5 \\
P22 & Attending & >10 \\
P23 & Attending & 5-10 \\
P24 & Attending & 3-5 \\ \bottomrule
    \end{tabular}
    \caption{Summary of study participants, their roles, and their level of experience working in the ICU. ICU = intensive care unit; APP = Advanced Practice Provider.}
    \label{tab:demographic_info}
\end{table}
% \begin{table}[]
%     \centering
%     \begin{tabular}{cc|ccc|ccc|ccccc}
%     \toprule
%     \multicolumn{2}{c}{Gender} & \multicolumn{3}{c}{Race and Ethnicity} & \multicolumn{3}{c}{Current Role} & \multicolumn{5}{c}{Years of ICU Experience} \\
%     \midrule
%     Male        & Female       & White    & Asian   & Hispanic/Latino   & APP    & Fellow    & Attending   & \textless 1   & 1-2   & 3-5  & 5-10  & 10+  \\
%     \midrule
%     18          & 6            & 17       & 6       & 3                 & 6      & 3         & 15          & 1             & 2     & 6    & 5     & 10  \\
%     \bottomrule
%     \end{tabular}
%     \caption{Summary of participant demographics}
%     \label{tab:demographic_info}
% \end{table}

During the study, participants used a simplified AI Clinician Explorer interface to assess and make treatment decisions for four patients while thinking aloud. Patients were selected from the MIMIC-IV dataset by the authors, as described in Sec. \ref{sec:case-selection}. Participants were free to explore all patient data prior to the time-step of interest, which included demographics, vital signs, lab values, mechanical ventilation settings, and a record of all treatments administered since the beginning of the patient's ICU stay.

\begin{figure*}
    \centering
    \includegraphics[width=\textwidth]{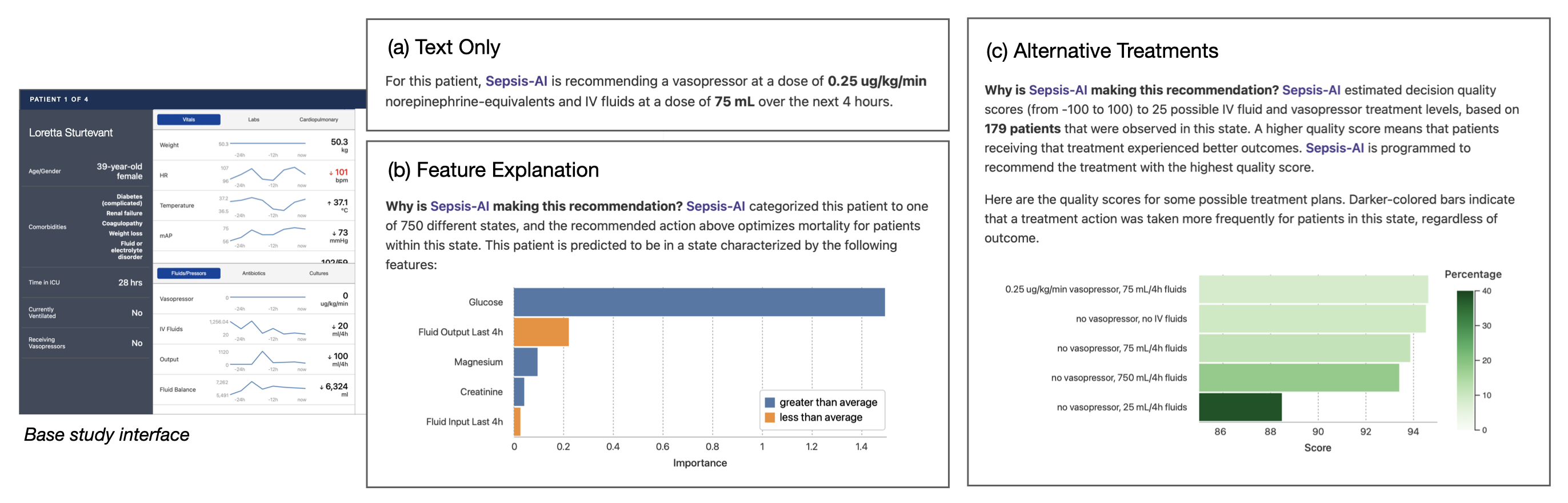}
    \caption{Visualization conditions used in the study. All participants were shown the base interface with patient trajectory views from the AI Clinician Explorer. The right half of the interface contained one of the following conditions when the AI was shown: (a) A textual description of the AI's recommendation, introduced in the Text Only condition and shown alongside subsequent visualization conditions as well. (b) The Feature Explanation chart shows the five variables that contributed most strongly to the AI's characterization of the patient state, and how each variable's values deviate from the average. (c) The Alternative Treatments chart shows five possible actions and the frequency at which clinicians historically took each action.}
    \label{fig:visualization-conditions}
\end{figure*}

Patients were presented in a randomized order. For each patient, participants saw a different version of the AI recommendation (``visualization condition''), presented in a fixed order with each successive condition containing more information. We elected to present visualization conditions in a fixed order to minimize cognitive burden on our participants and to give them an opportunity to progressively acquaint themselves with the features of the AI Clinician interface. As summarized in Fig. \ref{fig:visualization-conditions}, the visualization conditions were as follows:

\begin{enumerate}
    \item \textbf{No AI.} Participants made the decision without an AI recommendation.
    \item \textbf{Text Only.} Participants were introduced to the AI and given a simple text-based recommendation (e.g., ``For this patient, the AI recommends...'')
    \item \textbf{Feature Explanation.} In addition to the textual recommendation, participants were shown a SHAP feature attribution chart explaining how the patient's state was determined (Fig. \ref{fig:visualization-conditions}b).
    \item \textbf{Alternative Treatments.} Finally, for this condition participants were shown a bar chart with five possible treatment actions ranked by the AI-generated quality score. Bars were also color-coded by the frequency at which clinicians in the historical dataset took each action for similar patients, providing participants with a sense of both how common a decision was and the quantity of data that the recommendation was based on (Fig. \ref{fig:visualization-conditions}c)\footnote{Although this visualization includes two different types of information (AI-predicted value and aggregate clinician behavior), we opted to include it as a single experimental condition to minimize the study burden for participants while collecting relevant think-aloud feedback for future iterative design.}.
\end{enumerate}
The AI was introduced to participants as ``Sepsis-AI,'' a tool that ``analyzes patients’ electronic health records and uses an artificial intelligence-based algorithm to recommend fluids and vasopressor doses that optimize mortality based on historical data.'' To prevent bias we avoided referring to the AI as the ``AI Clinician,'' since some participants may have been familiar with the discussion surrounding the original publication.

After reviewing each patient's history and current status, participants were asked to choose a treatment action to apply to the patient related to both the IV fluid amount and vasopressor dosage. Although the recommendation included specific dosages of IV fluids and vasopressors, we limited participants' choice set to (up to) three options in an effort to capture clinicians' first-order decision-making process,  and to better replicate the way clinicians make resuscitation decisions at the bedside \cite{mansoori2022impact}. This design had the added benefit of increasing the analytic tractability of our results. The three options were: begin/increase, end/decrease, or leave unchanged. If a treatment strategy was not currently being used (e.g. patient not on vasopressor), the end/decrease option was removed, leaving participants with 4-6 possible actions per patient. After making each treatment decision, participants reported their confidence in their own treatment choice (on a 7-point Likert scale bounded by ``not at all confident'' and ``extremely confident'') and their beliefs about how challenging the case was (on a 7-point Likert scale bounded by ``extremely easy'' and ``extremely challenging''). For all visualization conditions except for No AI, participants also rated the usefulness of the Sepsis-AI recommendation (on a 7-point Likert scale bounded by ``not at all useful'' and ``extremely useful'') and the degree to which the Sepsis-AI recommendation affected their confidence in their own treatment choice (on a 7-point Likert scale bounded by ``... much less confident'' and ``... much more confident'').

Once participants had entered their decisions on all four cases, we concluded the session with a brief semi-structured interview to understand how clinicians used the different visualizations as well as their perspectives on when they might perceive the AI to be helpful.

\subsection{Case Selection}\label{sec:case-selection}

In any study involving acceptance of AI-generated recommendations, the scenarios that are chosen can have a large impact on participants' level of concordance with, and perceived trust in, the AI. As discussed in Sec. \ref{sec:model-description}, a variety of factors make the ``accuracy'' of the AI Clinician's treatment recommendations impossible to determine with certainty. Therefore, instead of choosing cases based on a target level of accuracy, we deliberately chose cases and decision points in which the AI Clinician's recommendation was substantially \textit{different} from historical clinician actions. Specifically, we used the AI Clinician Explorer to identify patients (and timesteps within patients) in which the AI Clinician recommended one treatment strategy (e.g. vasopressors and no IV fluids) for patients in a particular state, but a plurality of clinicians gave an alternative treatment (e.g. IV fluids and no vasopressors). This approach had the added benefit of replicating situations in which the AI recommendations might challenge clinician judgment. To make the cases more realistic, each patient was given a randomly-generated name, and the visualization was accompanied by a hypothetical clinical vignette summarizing the patient's status. The vignettes were written to provide only generic clinical context (e.g. ``sepsis from a urinary tract infection''), with no information that could guide treatment decisions beyond what was included in the dataset. A summary of the cases is shown in Table \ref{tab:patient-cases}.

\begin{table*} \small
\begin{tabular}{rp{3cm}p{3cm}p{3cm}p{3cm}}
\toprule
\textbf{Patient Pseudonym} & \textbf{Ruth Silva} & \textbf{Loretta Sturtevant} & \textbf{Jeffrey Williams} & \textbf{Victoria Thompson}                                                    \\ \midrule
Demographics                & 76 y/o female                                                      & 39 y/o female                                                               & 74 y/o male                                                                                             & 63 y/o female                                                        \\
Key Characteristics         & mechanically ventilated, undiagnosed sepsis, currently hypotensive & type I diabetes, chronic renal insufficiency, previously received IV fluids & congestive heart failure, mechanically ventilated, recent admission, currently on high dose vasopressor & previously received vasopressor and IV fluids, currently hypotensive \\

AI Recommendation           & \multilinecell{no change in fluids\\increase pressors} & \multilinecell{increase fluids\\increase pressors} & \multilinecell{increase fluids\\decrease pressors} & \multilinecell{increase fluids\\no change in pressors} \\

Original Clinician Decision & \multilinecell{increase fluids\\no change in pressors} & \multilinecell{increase fluids\\no change in pressors} & \multilinecell{increase fluids\\decrease pressors} & \multilinecell{no change in fluids\\increase pressors} \\

Majority Attending Decision & \multilinecell{increase fluids\\no change in pressors} & \multilinecell{increase fluids\\no change in pressors} & \multilinecell{increase fluids\\decrease pressors} & \multilinecell{increase fluids\\no change in pressors} \\ \bottomrule
\end{tabular}
\caption{Summary of the four patient cases selected for the think-aloud study. De-identified patient data was derived from the MIMIC-IV dataset. Three reference treatment decisions are shown for the time interval at which patients were presented: the AI Clinician's recommendation, the decision that was made by the clinician that treated the actual patient in the MIMIC-IV dataset, and the decision taken by the majority of attending physicians in our study in the No AI condition.}
\label{tab:patient-cases}
\end{table*}

\subsection{Analysis}
\label{sec:analysis}

Ratings of confidence and AI usefulness were compared quantitatively to assess participants' attitudes towards each of the visualization conditions. Using the Python \texttt{statsmodels} package\footnote{\url{https://www.statsmodels.org}}, ordinary least squares (OLS) regression models were fit to each 7-point Likert scale outcome using the visualization condition as the only predictor. Models controlling for the participants' role, gender, and years of experience yielded similar results. All models cluster standard errors at the respondent level using robust Huber-White estimators. For post hoc (pairwise) comparisons, we adjust for multiple tests using the Holm–Bonferroni method.

In the absence of a ground truth correct decision, treatment choices were evaluated in terms of their \textit{concordance} against three reference standards for each patient: (1) the AI Clinician's recommendation, (2) the action taken by the clinician(s) on the actual patient in the MIMIC-IV database, and (3) the majority action chosen by attending physician participants in the No AI visualization condition. The latter served as an approximation for the ``clinical consensus'' decision for each patient, although (as expected) variability was observed even within these experts' decisions. To understand the relationship between visualization condition and concordance, we used  logistic regression in a similar fashion as above.

% These coders met to discuss and resolve coding discrepancies for each interview for codes that fell below 80% agreement.

The 12.6 total hours of think-aloud sessions were machine transcribed using Descript\footnote{\url{https://www.descript.com}} and manually cleaned in preparation for qualitative analysis. After reviewing these transcripts and notes in an interpretation session, the team developed a set of 23 codes that could systematically capture distinct decision-making behaviors. The four segments corresponding to each patient case were excerpted from each transcript and coded using this code book by two members of the research team. These coders met to discuss and resolve coding discrepancies and refine code definitions as needed. Finally, participants' broader viewpoints on decision-making using the AI were extracted using open coding, and themes were identified using affinity diagramming.

\section{Results}

The following sections provide first an overview of participants' attitudes towards the AI in each of the visualization conditions (Sec. \ref{sec:quantitative-findings}), followed by the decision-making behavior patterns observed in the think-aloud transcripts that help explain participants' use of the AI (Sec. \ref{sec:decision-patterns}). We then describe how as expert decision-makers, participants interrogated the underlying assumptions of the AI we presented them with, and reflected on how it could better assist them (Sec. \ref{sec:ai-perspectives}).

\subsection{Perceptions of Decision-Making with AI and Explanations}\label{sec:quantitative-findings}

% The visualization conditions we presented had significant effects on participants' perceptions of the AI and of their own decision-making. Overall, the Feature Explanations condition was associated with higher perceived AI usefulness as well as stronger positive effects on confidence. However, participants' verbal feedback indicated that they found specific parts of the Alternative Treatments visualization useful as well. Finally, visualization conditions also affected participants' perceptions of case difficulty. Below we report quantitative findings for the four Likert-scale responses we measured, alongside relevant qualitative responses that help contextualize the data.

\begin{figure*}
    \centering
    \includegraphics[width=0.95\textwidth]{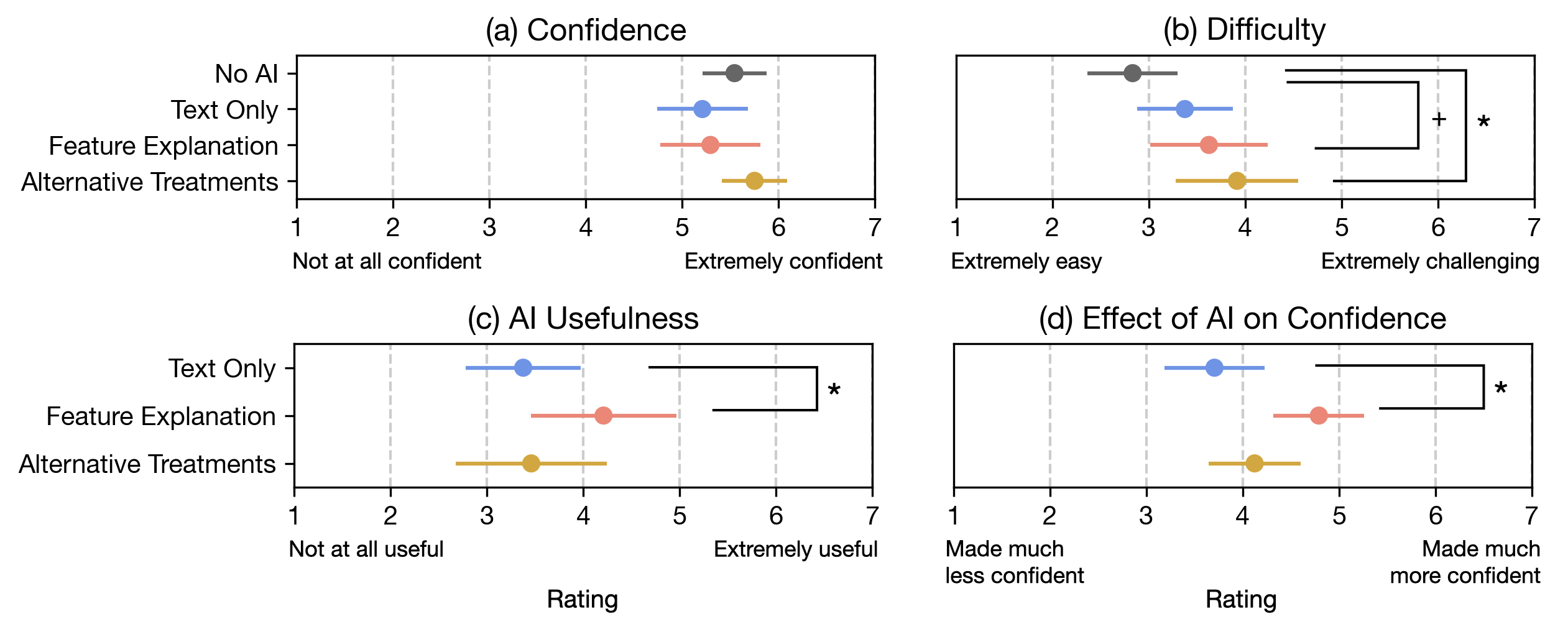}
    \caption{Summary of quantitative measures obtained from participants' self-ratings within each visualization condition: (a) participants' confidence in each decision; (b) how difficult they rated each case; (c) their rating of the usefulness of each version of the AI; and (d) how much the AI affected their confidence. * signifies $p < 0.05$; + signifies $p < 0.1$. Error bars indicate 95\% confidence intervals.}
    \label{fig:quantitative_results}
\end{figure*}

% An alternative paragraph to put here.
Across several measures, participants’ perceptions of the AI varied as a function of visualization condition. Participants reported that the AI was more useful and that it increased their confidence to a greater degree when participants saw one of the two \textit{explanation} conditions (Feature Explanation or Alternative Treatments), relative to when they saw the Text Only recommendation. Below, we report quantitative findings for the four Likert-scale responses we measured alongside relevant qualitative responses that help contextualize the data. The full pattern of results is reported in Fig. \ref{fig:quantitative_results}.

\textbf{Usefulness of the AI.}
Visualization condition was associated with significant differences in participants' ratings of the AI's usefulness ($F(2, 69) = 4.251$, $p = 0.03$). Participants rated the AI as being more useful in the Feature Explanation condition than in the Text Only condition ($\Delta = 0.83$, 95\% CI  $[0.24, 1.43]$, $p = 0.018$) and directionally more than in the Alternative Treatments condition ($\Delta = 0.75$, 95\% CI $[-0.03, 1.53]$, $p = 0.12$).

\textbf{Effect of AI on confidence.}
Similarly, Visualization condition affected how participants rated the AI's impact on their confidence ($F(2, 69) = 7.946$, $p = 0.002$). Participants reported that the AI had a more positive effect on their confidence in the Feature Explanation condition than in the Text Only condition ($\Delta = 1.08$, 95\% CI $[0.51, 1.66]$, $p < 0.001$) and directionally more than in the Alternative Treatments condition ($\Delta = 0.67$, 95\% CI $[-0.05, 1.38]$, $p = 0.13$).

In the think-aloud sessions, several participants mentioned the positive effects of seeing explanatory evidence, either in the form of the Feature Explanation or Alternative Treatments. In the latter condition, clinicians particularly appreciated the AI's ability to compare outcomes of multiple possible decisions (P16, P18, P24): \textit{``Seeing the different outcomes to those decisions in a similar case, I think is... the most convincing to change your clinical decision making''} (P24). However, the ability to see other clinicians' actions in this condition was less uniformly endorsed. Some respondents appreciated the additional reassurance of the sensibility of their decisions (P5, P7, P8, P24), while others (P10, P12, P17) expressed concern that it would steer novice clinicians towards common errors committed by less experienced clinicians: \textit{``I'm highly suspect of what other people do. And I don't think that that's a good way to practice medicine''} (P17).

\textbf{Confidence in treatment choice.}
Participants' confidence in their treatment choices was not significantly different across Visualization conditions ($F(3, 92) = 2.220$, $p = 0.11$) and no pairwise comparisons between conditions were statistically meaningful after adjusting for multiple comparisons ($p\text{s} > 0.17$). However, there was a directional increase in confidence ratings when explanatory visualizations were provided, particularly in the Alternative Treatments condition. We hypothesize that one benefit of the Alternative Treatments condition on decision confidence may have been that it presented evidence for multiple treatment options, not just the often-discordant top recommendation.

\textbf{Perception of case difficulty.}
Visualization condition significantly affected perceptions of case difficulty ($F(3, 92) = 4.112$, $p = 0.02$), with the provision of AI and its associated explanations increasing perceived difficulty. Comparing individual conditions, we find that participants perceived the cases as being significantly less challenging in the No AI condition than in the Alternative Treatments condition ($\Delta = 1.08$, 95\% CI $[0.46, 1.71]$, $p = 0.003$) and directionally less challenging than in the Feature Explanation condition ($\Delta = 0.79$, 95\% CI $[0.14, 1.45]$, $p = 0.09$). Our interpretation of this pattern is that explanatory evidence may have prompted clinicians to consider more factors when making their decision, especially when explanations did not align with their mental model of the patient or when the recommendations went against their clinical judgment (P12, P17, P22, P24). The resulting cognitive burden may have made the case seem more difficult. For instance, one clinician noted:

\begin{quote}\textit{``I would not have guessed that the decision or the recommendation was being based on something like a BUN [blood urea nitrogen] change. I assumed it was based on the CVP [central venous pressure], and I don't think that CVP was considered in [the Feature Explanation chart]. And so it kind of makes you try and guess where the recommendations are coming from, and you spend a little bit more mental energy thinking about that.''} (P17)\end{quote}

\subsection{Patterns of Interaction with the AI}\label{sec:decision-patterns}

\begin{figure*}
    \centering
    \includegraphics[width=0.75\textwidth]{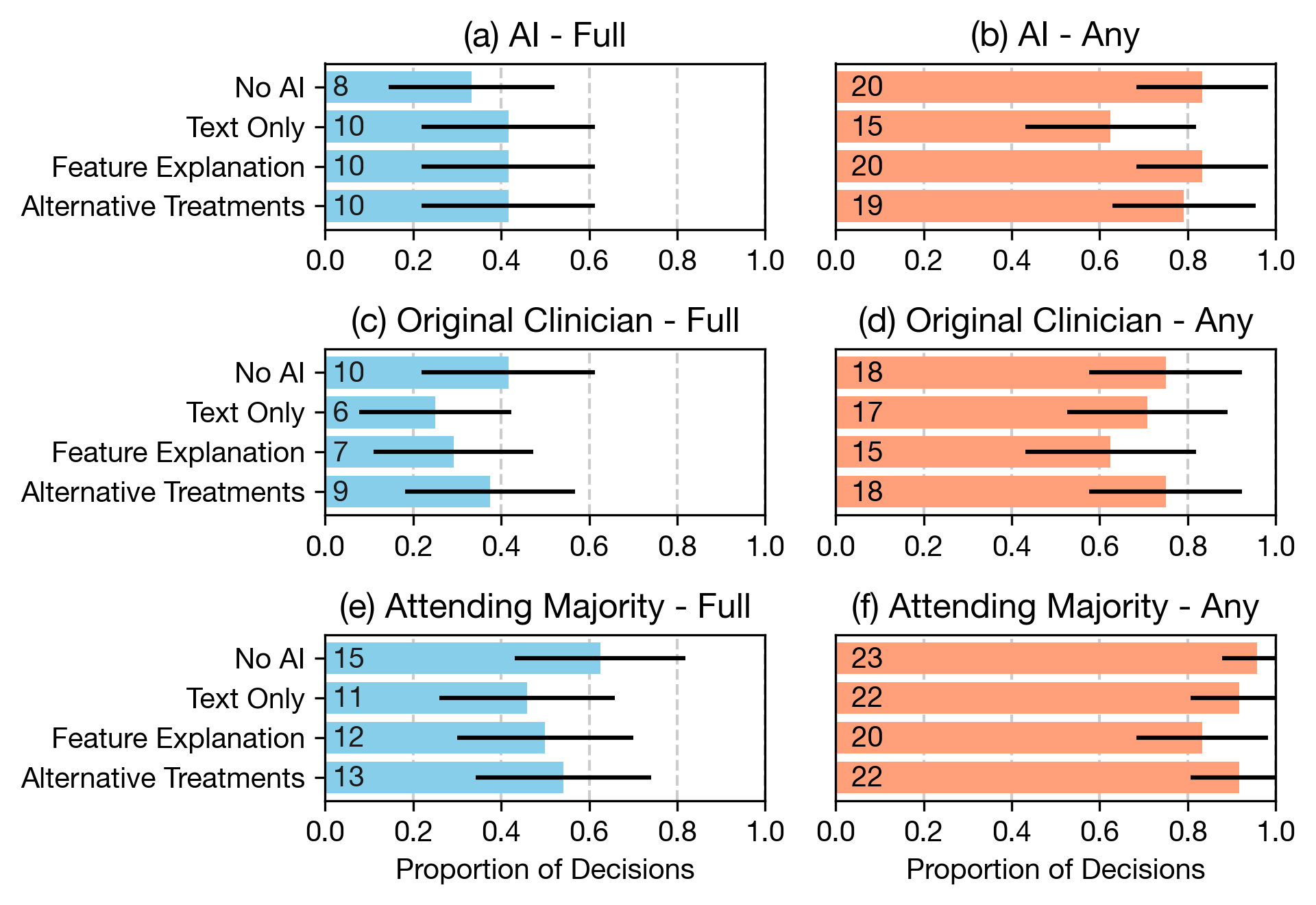}
    \caption{Rates of concordance between participants' decisions and three reference decisions: the AI recommendation, the decision of the clinician in the original dataset, and the decision taken by the majority of attending physicians in the No AI condition. The left column (``Full'') shows agreement with both the IV fluid and vasopressor recommendations, while the right column (``Any'') depicts agreement for \textit{either} of the two treatment strategies. Each proportion is calculated over a total of 24 decisions; error bars indicate 95\% confidence intervals.}
    \label{fig:ai-concordance}
\end{figure*}

In contrast to the attitudinal metrics, participants' actual decisions for each patient did not vary meaningfully as a function of visualization condition. The light blue bars in Fig. \ref{fig:ai-concordance}a show that clinicians chose the same treatment choice as the AI about 42\% of the time regardless of the visualization condition—only a slight increase over the 33\% base rate of concordance without seeing the recommendation at all. If any concordance (same choice according to \textit{either} fluids or vasopressors) is included, participants again have roughly similar rates of agreement with the AI, except for a slightly lower rate in the Text Only AI condition (Fig. \ref{fig:ai-concordance}b). 

When the AI was shown, we did observe a slight reduction in concordance with actions taken by the clinician treating the original patient as well as the majority attending decision (Fig. \ref{fig:ai-concordance}c-f). Specifically, the average full concordance with the majority attending decision was 50\% across the three AI conditions, compared to 63\% in the No AI condition. This may indicate that participants were swayed to do something other than the ``typical'' clinician action when using the AI. Yet the actions they ultimately took did not perfectly align with the AI either: out of the 36 AI-assisted decisions in which the participant did not fully agree with attendings, only 6 decisions showed full concordance with the AI. Though not statistically significant by logistic regression modeling, these somewhat counter-intuitive relationships led us to hypothesize that individual-level variations could be contributing to the roughly-constant overall rate of concordance with the AI. 

\begin{figure*}
    \centering
    \includegraphics[width=0.76\textwidth]{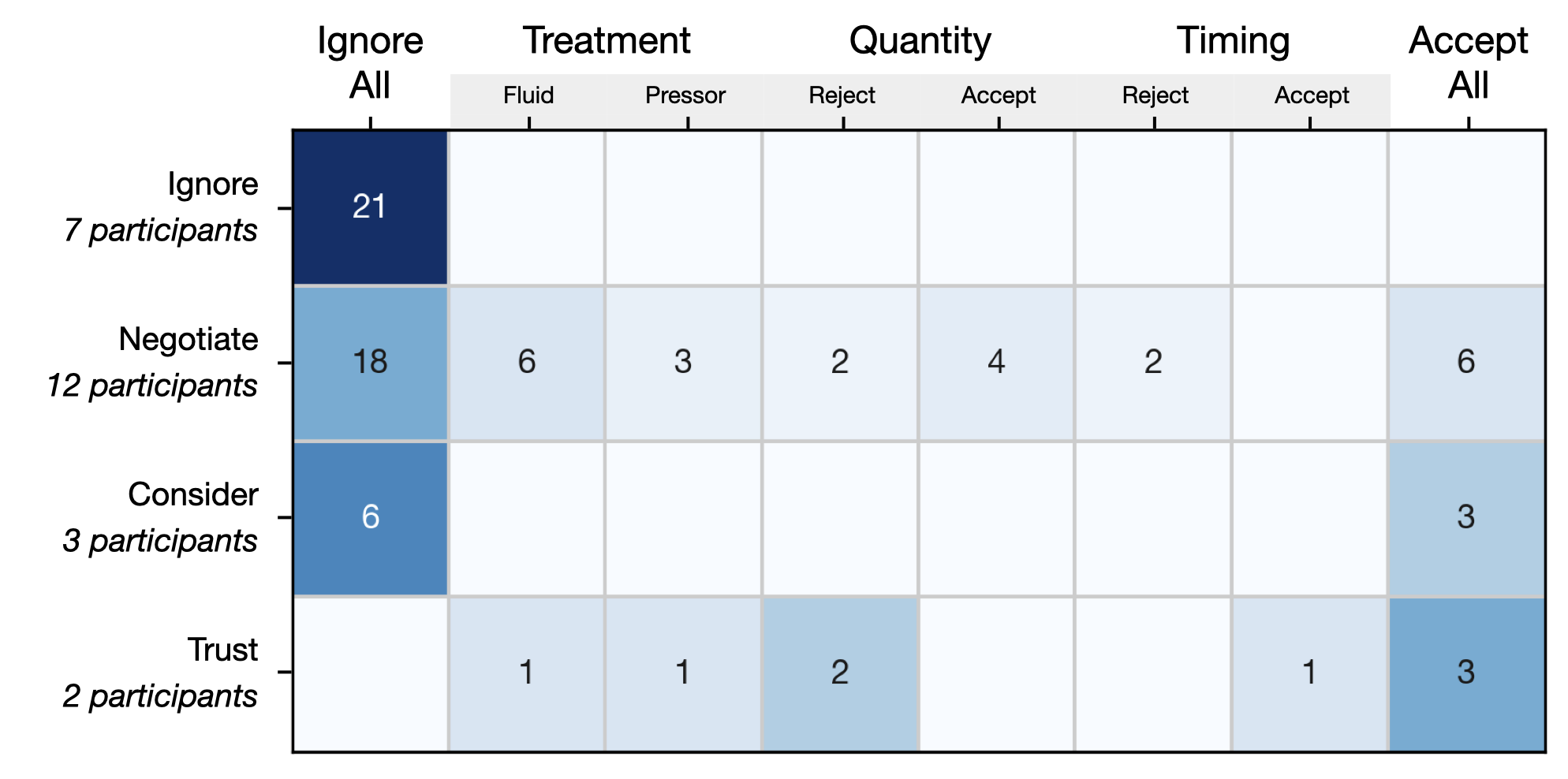}
    \caption{Patterns of reliance observed in participants' decisions, summarized from qualitative coding of think-aloud transcripts. The columns represent behaviors observed in an individual decision: Ignore All (decision not affected by AI), Treatment (accepted recommendation on one treatment strategy but not the other), Quantity (dosage levels), Timing (when to administer each treatment), and Accept All (fully changing the decision to align with the AI). The rows are sets of participants, grouped by these behaviors. Colors are normalized within participant groups (3 decisions per participant using the AI).}
    \label{fig:reliance-behaviors}
\end{figure*}

Therefore, to gain more granular insight into when participants chose to accept AI recommendations, we turned to the qualitative analysis of participants' think-aloud transcripts. As described in Sec. \ref{sec:analysis}, we developed codes to capture whether and how participants engaged with the AI along various aspects of its recommendations, as well as the reasons they provided for accepting or rejecting the recommendation. Grouping together participants with similar codes revealed four distinct behavior patterns, each of which was associated with different degrees of reliance on the AI. The four behavior patterns are summarized in Fig. \ref{fig:reliance-behaviors} and described in more detail below.

\subsubsection{Ignore: Participant makes own decisions}\label{sec:ignore-group}

For seven participants (21 total decisions using the AI), the AI never meaningfully influenced their decision in any way indicated by their think-aloud transcripts. Instead, their decision was predominantly driven by their initial clinical assessment, and not affected by recommendations even when explanatory visualizations were provided. These participants were often able to reject the recommendation because they were highly confident in their decision already, due to characteristics of the patient they identified as important based on their clinical experience: \textit{``She's young and doesn't have heart problems and she's very net negative. So fluids would be the first thing I do for her, for sure''} (P5). Perhaps as a result of their confidence, these participants sometimes gave no verbal acknowledgement of the AI recommendation (3/21 decisions) despite the fact that it was clearly demarcated to them and they knew it was present. When participants did engage with the recommendation, they tended to critique it while holding their own assessment fixed. For example, P11 rejected a recommendation to give vasopressor and a small amount of IV fluid, arguing:
\begin{quote}
    \textit{``She may be hypovolemic... because of the hyperglycemic state, but certainly I would not... start a pressor on this patient. [...] And IV fluids at a dose of 75 mLs over the next four hours... I disagree with that as well, because I think that this patient might be losing a lot of fluid on the urine output because of hyperglycemia.} (P11)
\end{quote}

Interestingly, this engagement with the recommendation also affected some participants' confidence despite not affecting their decision. In these cases the AI served to either confirm the initial assessment—\textit{``made me feel better about that decision'' (P5)}—or, more commonly, to induce doubt when the recommendation was discordant (P6, P7, P17, P18). For instance, P17 noted that the recommendation \textit{``to a certain degree made me question more than I would've. It actually probably made me think more about starting vasopressors, when any other time I would've just given the fluid bolus and not thought about it.''} However, because these participants were already confident in their clinical reasoning and fairly settled on their decision, the AI recommendation was insufficient to cause them to change course.

%\subsubsection{Discomfort at discordance with the AI}

\subsubsection{Negotiate: Participant chooses aspects of the recommendation to accept}\label{sec:negotiate-group}

Unexpectedly, the most common behavior pattern we observed was of participants selectively adopting aspects of the recommendation as a form of auxiliary evidence. As with the Ignore group, the twelve participants in the Negotiate group still frequently made decisions that were not influenced by the AI (18/36 decisions). But in many cases, as shown in the middle columns of Fig. \ref{fig:reliance-behaviors}, they accepted at least one aspect of the recommendation:

\begin{enumerate}
    \item \textbf{Overall treatment choice.} In 14/36 decisions, participants agreed with the treatment recommendation for either fluids or vasopressors, but not both. For instance, P6 initially decided to follow a recommendation to begin vasopressors; however, upon re-examining the patient data, they noticed: \textit{``She hasn't gotten any [fluids]... okay, interesting. Hmm. I would probably give a little fluid too.''}
    
    \item \textbf{Quantity of treatment.} Participants engaged with the AI recommendation's specific dosage levels in 12/36 decisions, most often rejecting the values based on their knowledge of the patient: \textit{``She's 39. I know she has chronic renal failure, but that doesn't mean that she cannot use fluid''} (P10). However, when the dosage values were within the range that participants would expect, they found value in the specificity of the AI recommendations: % You very rarely give 25 mLs an hour of fluid maintenance to anybody, particularly when they're septic, hypotensive, and looking like this.
    \begin{quote}
        \textit{``I think a big challenge in the ICU is having a sense of... how much fluid to give a patient. [...] I think in that situation, I'm sort of more willing to give [the AI]... more of the nuance of the decision making. Like the big picture, we both seem to be in agreement. [...] And so then if the AI says, `this is how much fluid I think they need in this period of time,' that's one less decision that I have to tax myself or burden myself with.''} (P8)
    \end{quote}
    
    \item \textbf{Timing of treatments.} Finally, some participants expressed agreement with the AI's overall recommendation but refrained from making the recommended changes concurrently. For instance, P7 deferred the vasopressor component of one recommendation, reflecting that \textit{``I don't necessarily disagree, it's a relatively small dose of norepinephrine. [...] I would probably start with the fluids, but then I would escalate to vasopressors if there was no response probably within a couple hours.''} Conversely, one participant was swayed by a vasopressor recommendation to postpone their own decision to give fluids (P22).
    % \begin{quote}
    % \textit{``So I would begin vasopressors. I think that's right... And if it didn't [improve blood pressure], then I would probably give a fluid challenge. And I concede that having the AI suggest what it suggested may have steered me away from giving fluids in addition to the vasopressors.''} (P22)
    % \end{quote}
    % These participants often contrasted the timing of the recommendation with their experience and standard ``order of operations'' of treating sepsis. However, they interpreted the AI's recommendation as a form of evidence that could be integrated into their current decision, allowing them to apply the optimal treatment faster.
\end{enumerate}

We termed this behavior \textit{negotiation} because participants assigned value or priority to various aspects of the recommendation, and thereby were able to arrive at an intermediate solution that balanced its most important aspects with their own intuition. Participants often prioritized parts of the recommendation using two factors:

\begin{enumerate}
    \item \textbf{Risk level and urgency.} In 12/36 decisions, participants used their perception of the severity of the patient's sepsis to decide how much to reconsider their treatment plan. Discordant recommendations for patients whose vitals seemed relatively stable were more likely to gain acceptance than those that appeared to be deteriorating. For example, weighing a recommendation to give vasopressors against their initial assessment to give fluids, P3 responded, \textit{``I would say if I was alone without the computer helping me, I would give a trial of fluid. But I'm comfortable doing what they say. I think it's a coin toss anyway.''}
    \item \textbf{Evidence presented by the AI.} More so than other groups, participants in this group used the explanatory visualizations as a source of evidence with which to understand the main point of the AI recommendation. For example, P18 was convinced by the Alternative Treatments chart to give fluids over their initial decision to start vasopressors:
    \begin{quote}
        \textit{``Looks like very few people would have gone back on pressors, which is what I wanted to do. [...] I think it's fair. She needs something modestly aggressive because her [blood pressure] is quite low and it's been falling. [...] Yeah, I think this is a reasonable choice.''}
    \end{quote}    
    Interestingly, this participant was discouraged from taking a less-common path by not only the AI's recommendation, but the summary of aggregate clinician behavior that the Alternative Treatments chart provided. On the other hand, negotiations sometimes led clinicians to ultimately reject the recommendation because they could not justify to themselves how an explanatory chart led to the recommendation. For instance, on a Feature Explanation chart, P24 questioned \textit{``why a low [blood urea nitrogen] would lead to starting fluids and not vasopressors,''} ultimately leading them to go against the AI.
\end{enumerate}

Perhaps because of the additional value they were able to obtain from the AI, Negotiate participants rated the recommendations more useful than other groups, with an average 7-point Likert rating of 4.6 ($SD = 1.27$) compared to 2.8 ($SD = 1.79$).

\subsubsection{Consider: Participant conditionally accepts or ignores the recommendation}\label{sec:consider-group}

Three participants were similarly open to accepting the AI recommendation as the Negotiate group, but they either fully relied on the AI or made the decision on their own. Specifically, in 3/9 of their decisions, they yielded control of the decision to the AI, primarily based on their sense of uncertainty. For instance, P9 resolved to follow the AI recommendation for a difficult case: \textit{``I am ambivalent about this one. Her [blood pressure] is slightly low. Her heart rate is actually coming down, fluid balance is positive... I think it's fine. We can do what the AI recommends.''} Conversely, the same participant confidently dismissed a different recommendation: \textit{``For this patient [the AI] is recommending a vasopressor dose of 0.25 of norepinephrine? Yeah, I don't think so.''} In this way, the three Consider participants used the AI to drive their decisions when they were uncertain, but resumed control of decision-making in highly certain cases.

\subsubsection{Trust: Participant always accepts some part of the recommendation}\label{sec:trust-group}
Finally, two participants were influenced by the AI for at least part of their decision in \textit{every} decision they made. These participants often emphasized that the AI was based on objective data, perhaps leading them to consider its recommendations more willingly than other participants. For example, after reviewing the Alternative Treatments visualization, P16 reflected, \textit{``This higher score means that they had better outcomes? Well then I'm gonna have to go with that. [...] The data looked pretty good.''}

\subsection{Perspectives on AI for Treatment Decision-Making}\label{sec:ai-perspectives}

Throughout and after the think-aloud portion of each session, participants commented on how their decision-making processes in the simulated study environment compared with the decisions they made on a day-to-day basis. They also reflected on their habits and standard practices as clinicians, and how an AI might or might not be used to beneficially transform those practices. Below we discuss four themes that emerged from these discussions.

\subsubsection{Participants' decisions are often guided by bedside information-gathering techniques rather than metrics used by the AI}
We did not explicitly probe for next actions other than IV fluids and vasopressors, but eleven participants mentioned that a helpful next step would be additional data collection in the form of bedside assessment unavailable to the AI system. This usually took the form of dynamic assessments for fluid responsiveness via the physical exam, a procedure known as a ``straight leg raise,'' or use of bedside ultrasound imaging. For instance, P21 noted that this information could help resolve a conflict with the AI on how much IV fluid to administer: \textit{``If the AI was disagreeing with me, what I would do is walk into the room, do a leg raise, do a ultrasound... and then based on that information, I would decide how much volume to give.''} In fact, participants viewed this information as more reliable than any data used by the AI. They used this distinction to assert the superiority of human decision-making, reinforcing their identity as expert decision-makers while not outright rejecting the AI recommendation:
\begin{quote}
    \textit{``At the bedside, I would acquire one piece or two pieces of reliable, better quality data than the algorithm has available. And then I would use that to make my decision [...] It's not fair to ask an algorithm to make a prediction that is as reliable as that is, because it doesn't have access to that.''} (P23)
\end{quote}

Participants similarly expressed concerns that the AI did not have access to more gestalt characteristics such as the patient's general appearance (P3, P7, P13): \textit{``How ill do they look?''} To be clear, participants could not use these assessments during the study either, as they could only view the numerical data and general patient vignettes that we provided. Nevertheless, some clinicians (P20, P23) contrasted their confidence in these contextual assessments against the statistical nature of the AI: \textit{``My bias as a clinician is that there is significant between-patient variability that is clinically significant, such that population level estimates used to inform individual patient care is fraught''} (P23).

\subsubsection{The discretized dosage levels and time-scales used by the AI do not match with clinical practice}

% Another perceived source of disconnect between the AI and participants' decision-making was the amounts and frequencies of treatments that the AI calculates in order to systematize evaluation over thousands of trajectories. 
By design the AI Clinician collapsed all fluid and vasopressor dosage levels into 25 bins based on quantiles, ensuring a roughly uniform distribution of training labels. However, in practice this discretization led to confusion and doubt because all of the IV fluid bins were relatively low compared to the amounts clinicians were used to (presumably because most timesteps did not involve substantial fluid administration). For example, the third treatment level for fluids is 75 mL over four hours, to which one participant commented, \textit{``I've never ordered such a small dose of fluids... To me that's like sprinkling water on her''} (P18).

The AI also aggregates data and provides recommendations at 4-hour intervals, which balances the rate of biometric data availability in the training dataset with the typical frequency of decision-making in the ICU. Clinicians overall found the 4-hour timescale appropriate for viewing the patient's trajectory and for making decisions on relatively stable patients, but they noted that they \textit{``would not feel comfortable''} committing to higher-risk treatment decisions over that duration (P4, P7, P14, P17). Shorter-term decision points were viewed as a buffer against uncertainty about treatment responsiveness: \textit{``In those situations where you're on the fence... you're gonna give your [IV fluid bolus], and you're gonna follow in that hour to two hours after they get the bolus to see if it had an effect''} (P14). In terms of measuring reliance on the AI, this reduction in timescale resolution led to clinicians effectively postponing agreement with the recommendation to a later decision (P4, P10, P17, P20), potentially nullifying the potential benefit of advance prediction by the AI.

\subsubsection{Clinicians become skeptical of AI when it deviates from standardized or individual care practices}

Participants often compared the AI's recommendations to the guideline-recommended practice of treating septic patients with hypotension, which comprises administering IV fluids (typically around 30 mLs per kilogram of body weight) and then vasopressors if the patient's blood pressure does not normalize \cite{Evans2021}. These guidelines explicitly state that there is room for variation and that individual treatment plans should still be customized to each patient's unique circumstances, a fact acknowledged by participants—\textit{``you have to sort of be willing to be flexible'' (P16)}. Nevertheless, eight clinicians mentioned during decision-making that they would expect the AI to recapitulate rather than deviate from the guidelines. For instance, one participant voiced the tension they felt between the AI's recommendation and their training: \textit{``So I see the score, but going off of the data and all of my knowledge of sepsis, we have to try to give her some fluids. We never jump straight to vasopressors''} (P19). It is impossible to know if the AI's recommendation to give vasopressors was a better decision, although some evidence shows that early administration of vasopressors could benefit patients \cite{Shi2020} and expert opinion increasingly emphasizes vasopressors over fluid administration \cite{Jozwiak2018}. Yet these recommendations were dismissed as nonsensical given the patient's current status: \textit{``Thinking she's not hypotensive. So why in the world is the AI asking me to start pressors? I'm rapidly losing faith in Sepsis AI''} (P12).

Participants also wanted the AI to concur with their personal practices, which they often defined in contrast to the predominant habits of other clinicians. For instance, two participants found the AI's recommendations \textit{``a bit fluid aggressive''} (P2) at times, particularly because they perceived that many clinicians overuse fluids: \textit{``I've seen it in ICU where we're just like bolusing them blindly. And the next thing you know, they're puffy like the Michelin man''} (P1). Five participants (particularly more experienced clinicians) framed their personal practices as the standard of comparison for both the AI and other clinicians, in that when \textit{``the recommendation starts not very in line with what I would personally do with the patient, I don't think it's useful''} (P20). Because they viewed the AI as based on the actions of a general population of clinicians less skilled than themselves, participants were able to dismiss recommendations that aligned with norms they were already comfortable deviating from.

\subsubsection{Rigorous and credible evidence of the AI's effect on outcomes is a prerequisite to trust.}

Aside from their reactions to individual decisions, several participants expressed that their overall level of trust in the AI would be determined based on the description of methodology and evidence provided to them before they ever used the tool (P3, P9, P12, P17). These participants believed they would read available background information on the tool, then either \textit{``adopt it as a valuable tool or... shoot holes in it and say, `I don't believe in this methodology and I'm not gonna use this tool anyway'''} (P17). The credibility of the AI would partially be determined by the reputability of its developers and the journal in which its validation study was published: \textit{``If... there was a study in New England [Journal of Medicine] that said that Sepsis AI... improved outcomes, then I would say it could be kind of useful''} (P9). Once a high volume of credible evidence was available in favor of using the AI, participants believed they would more willingly trust its recommendations (P3, P9).

Although participants agreed that rigorous and credible evaluation was required, they were divided on how such a tool should be evaluated. The most common suggestion was to conduct a randomized controlled trial with the AI to validate whether the second opinion it provided improved patient care (P12, P23); others suggested simply testing the association between recommendation acceptance and patient outcomes (P7, P9). In contrast, P17 suggested that the AI should simply use their decisions as the ground truth and aim to replicate them, as is currently done for diagnostic models:
\begin{quote}
    \textit{``You could be convinced if somebody presented this to you and said, `Hey, we've been looking at your clinical practice, and... you're 95\% aligned with this. And so, you know, if we just set this to run, it's going to do the same thing that you would do 95\% of the time, and you don't have to wake up.''' (P17)}
\end{quote}
Regardless of what form the validation study took, participants agreed that upfront knowledge about the model's quality would not supersede clinical judgment on individual cases, leaving the door open to patterns of conditional and partial reliance even after trust is established.

\section{Discussion}

We describe the development of an interactive CDS system for sepsis treatment, as well as a mixed-methods study that examined how clinicians interacted with that system to identify critical barriers to AI adoption in health care. Our results confirm prior findings suggesting that providing clinicians with explanatory evidence, either in the form of feature explanations or alternative treatment comparisons, can increase clinicians' perceptions of the AI's usefulness and confidence in their decisions \cite{Alam2021,Tschandl2020}. In terms of reliance on the AI, prior work studying reliance on CDS tools \cite{Bussone2015} and explainable AI \cite{Zhang2020,Wang2020} led us to expect that clinicians would calibrate their own certainty against the AI and make a binary decision about whether to accept its advice in each case. However, only a few participants (the Consider group, Sec. \ref{sec:consider-group}) exhibited this dichotomous form of reliance. Instead, most participants engaged in a more nuanced form of partial reliance on the AI, often involving a negotiation between the initial clinical assessment and various aspects of the recommendation. Furthermore, several participants did not integrate the AI into their decision-making in any material way—for these participants the CDS only served to lower their confidence in their decision-making and increase the perceived difficulty of the case. Below, we discuss the implications of our results (key implications in bold) for the design of AI-based CDS and how to validate these systems in practice.

\subsection{Designing AI for Complex Clinical Decisions}

For a large number of health care decisions there is no evident ``right answer'' \cite{Eddy1984}. In these situations, successful AI should support clinicians in making better decisions on average, but must do so absent immediate feedback about the appropriateness of the recommendations. Out of the four broad decision-making behaviors we observed, the Negotiate behavior is closest to what one might consider an ``appropriate'' form of reliance on the AI in this setting. In contrast to the other three groups, participants who negotiated partial forms of reliance perceived a \textit{range} of plausible next steps for each patient, not just a single action stemming from their clinical assessment. Furthermore, they were able to override aspects of the recommendation when they had specific contextual reasons to do so. On the other hand, clinicians had to develop their own assessments of which parts of the recommendation to rely on, perhaps resulting in more inconsistent decisions.

\textbf{One approach to improve AI-assisted clinical decision-making could be to support negotiation by helping clinicians prioritize credible aspects of the recommendation.} For instance, instead of recommending a rigid treatment plan over a four-hour interval, an algorithm could leverage historical data to compare the value of starting multiple treatments concurrently with the value of applying them sequentially, helping inform comparisons that clinicians may already be making. Alternatively, it could present evidence in favor of general treatment strategies at a binary level (e.g. fluids and no vasopressors) rather than specific values (e.g. 250 mL of fluids) unless the specific dosage was known to have an impact on mortality. These systems would serve to reinforce the belief that humans can make more nuanced decisions than AI systems, a belief we observed in this study. This type of AI would ``know its limits'' but still be able to guide decision-making by providing a framework by which clinicians could inform their decisions, rather than providing only prescriptive recommendations that are easily rejected. Though technically non-trivial to develop, such an AI may yield advice that can be more easily and consistently assessed by clinicians.

Our study also examined the effects of model explainability, an ongoing area of debate in AI-based CDS research \cite{Amann2022,ArbelaezOssa2022,Ghassemi2021}, on participants' perceptions and behaviors using the AI. Our findings are consistent with prior XAI research \cite{Zhang2020,Wang2020,Bussone2015} showing that explanations are a helpful complement to AI predictions, but that explanations alone will not significantly impact reliance. In particular, we observed that while the Feature Explanation chart helped participants decide how much weight to place on the recommendation overall, it did not support their ability to assign value to individual recommendation components. On the other hand, the Alternative Treatments approach may have better supported negotiation behaviors by allowing the AI to \textit{``present its findings''} (P4) across a range of options. While similar to multi-class prediction charts used in prior work on diagnosis models \cite{Jacobs2021psych,Tschandl2020}, the fact that the actions depicted in our visualization were quantitative (i.e., specific dosage levels) may have yielded the additional benefit of helping participants understand the overall trend predicted by the model, and thereby negotiate intermediate solutions. \textbf{Future explainable treatment recommendation systems could extend the Alternative Treatments approach to facilitate more nuanced comparisons of different choices, such as by projecting future patient states and outcomes conditioned on different choice sequences.}

Another important finding for explainability is its potential effect on cognitive effort. AI is meant to improve the efficiency of clinical decision making, saving clinicians time and reducing workload. Yet we found that explainable AI has complex effects on cognitive effort, especially when clinicians must decompose every recommendation into aspects with differing levels of credibility. On one hand, participants in the Ignore group tended to lose confidence and waste time comprehending a recommendation that ultimately would not affect their choice. On the other, the visually dense explanations may have served as a cognitive forcing function to consider previously-neglected options \cite{Bucinca2021}, as Negotiate participants sometimes did. These results might suggest that the visibility and complexity of the AI recommendations be adjusted based on the users' confidence or the discordance between their decision and the AI. However, many participants also believed that once the AI was trusted, they would want to review it for confirmation of all their decisions, echoing Kulesza et al.'s findings that complete explanations tend to help despite requiring increased cognitive effort \cite{Kulesza2013}. \textbf{Further research is needed to understand the tradeoffs between providing confirming recommendations to build trust, and saving clinician effort on discordant but non-useful recommendations.}

One simple solution to improve AI acceptance could be to focus adoption efforts on novice clinicians that may lack confidence in their ability to independently make clinical decisions. However, contrary to prior work showing negative effects of task expertise and AI familiarity on acceptance of AI recommendations \cite{Gaube2021,Ehsan2021who,Jacobs2021psych,Bayer2021}, the behavior patterns we observed did not appear correlated to seniority or experience level. Two of the seven Ignore participants were not attending physicians, while both of the Trust participants \textit{were} attendings. While some prior work has examined how clinician demographics affect their needs for adopting AI \cite{Calisto2022}, our interviews suggested an additional factor to consider: many clinicians are already regularly exposed to decision rules and behavioral interventions derived from historical data and expert committees, and they often hold diverging beliefs about how this clinical advice should influence decision-making. Even with similar experience levels, clinicians express varying degrees of awareness (and skepticism) of how recommendations are generated \cite{Khairat2018,Yang2019}, but the effects of differences in these attitudes have yet to be examined. \textbf{A better understanding of these perspective differences, and how they relate to experience level, may lead to designs that better serve people reluctant to factor AI advice into their decisions.}

    % 1. Reshaping AI as a source of evidence rather than as a predictive agent
    % 2. Minimizing frequency that AI is used, providing recommendations contingent on non-optimal behavior
    % 3. Designing recommendations that support partial agreement
    
\subsection{Validating that AI-Based Decision Support Improves Outcomes}

Unlike much prior work on AI-assisted decision-making in health care \cite{Jacobs2021psych,Tonekaboni2019,Cai2019,Yang2019}, this study (1) used a real model trained to optimize treatment decisions, (2) provided clinical experts with real de-identified patient data, and (3) utilized a think-aloud protocol to capture further nuance beyond a multiple-choice survey. Participants responded to this realism in turn by revealing a more complex picture of clinical decision-making with an AI, one that in many ways does not fit the structure imposed by the AI. They expressed treatment goals in terms of information gathering (rather than always focusing on outcomes), adjusted dosage levels based on the patient's perceived needs, and temporally rearranged parts of the recommendation to more closely align with their standard practices. While this flexibility may well be desirable and even necessary in real-world decision-making, it creates an inherent tension with attempts to measure the quality of a system: \textbf{the more realistically an AI tool is integrated into clinical decision-making, the harder it becomes to assess whether the tool improves outcomes using standard validation techniques.}

Yang et al. \cite{Yang2019} and Amann et al. \cite{Amann2022} described a ``chicken-and-egg'' problem in which clinicians will not adopt AI recommendations unless they are backed by a credible validation study—yet in order for a validation study to succeed, clinicians need to adopt the AI's recommendations. This is particularly important in light of developing policies on AI in health care, such as the recent guidance by the U.S. Food and Drug Administration that treatment decision support systems such as the AI Clinician should be regulated as medical devices \cite{FDA2022}. But unless clinicians are obligated to use the AI as part of a randomized controlled trial, the AI's effectiveness in prospective validation will be confounded with clinicians' low level of trust in the system, resulting in a poor (and possibly over-optimistic) estimate of its performance in deployment. Compounding this challenge, our results suggest that binary acceptance or rejection of recommendations in the sepsis treatment context is not an accurate indicator of the AI's effect on decision-making. After all, participants often gave credence to the AI, yet they rarely followed its recommendation completely. In an \textit{in situ} validation study, how would partial or delayed acceptance be measured and assessed? \textbf{Developing acceptance metrics that account for partial reliance behaviors or changing reliance over time may help investigators perform validation studies that better capture potentially beneficial effects of the AI beyond binary acceptance.}

Another source of complication in validating AI-based recommendation systems is that reliability may vary significantly across different patient subgroups, requiring the user to develop a mental model of the AI's error boundaries \cite{Bansal2019}. However, even when clinicians in our study negotiated with the AI, they tended to approach its recommendations with a fixed level of trust or skepticism; their level of credence was rarely affected by the type of patient they were treating. \textbf{We suggest that instead of counting on end users to develop mental models of the AI's reliability, AI developers can collaborate with domain experts to extract, deploy and validate specific AI behaviors.} In other words, rather than considering the AI as an \textit{agent} whose advice needs to be evaluated across a wide range of clinical decisions, we propose to use AI as a \textit{source of evidence} whose recommendations can be separately assessed for specific subtypes of patients and disease states. This type of human-AI collaborative process could still yield more individualized recommendations than clinical trials (which are often too costly to run for all patient groups of interest), yet it would be more straightforward to evaluate than an AI that attempted to optimize for \textit{all} patients. These selectively-validated recommendations can then be introduced to clinicians in stages, building the credibility of the AI while minimizing the chance of unforeseen AI errors.

    % 1. Develop validation metrics that support partial or faceted reliance
    % 2. Validate specific AI behaviors (conditioned on specific populations) instead of the entire AI
    % 3. validation under time-evolving conditions - what if the same recommendations are shown later for less urgent cases?

\subsection{Study Limitations}

Although showing participants real AI recommendations for real patients yielded a more nuanced picture of decision-making, it also may have skewed our observations toward the particularities of the cases and recommendations we selected. Our depictions of the patient cases were limited to the structured data available in the MIMIC-IV dataset, meaning they had access to roughly the same amount of information as the AI. Additionally, we were unable to incorporate more domain-specific explanation techniques, such as explainable RL (XRL), since they would have required substantial changes to the previously validated AI Clinician model. As a result, the SHAP explanations we showed focused on only one part of the model (the state clustering), thus limiting their potential usefulness to end users. Future work should investigate whether using more transparent model architectures and RL-specific explanation strategies improves clinical utility over the visualizations we tested.

Our study design and recruiting strategy was primarily focused on obtaining a rich set of think-aloud data for every decision we observed. While this resulted in ample data for qualitative analysis, it also meant we were unable to assess the statistical significance of some of our quantitative results, particularly levels of concordance with the AI. In the future we plan to build on these results by conducting a similar study with a larger pool of participants, enabling us to more accurately estimate the effects of providing AI explanations. Importantly, the present work indicated a need for more granular ways to collect structured data about decisions, which will inform the design of subsequent survey instruments.

Finally, this study was conducted with clinicians at a renowned academic hospital system in the United States. As such, they were likely more familiar than the modal clinician with the idea of applying clinical protocols or AI tools to improve decision-making. However, it is not clear whether this familiarity would tend to make them more or less accepting of tools such as the AI Clinician. Further research in institutions that have been slower to adopt clinical decision support tools is needed to evaluate the generalizability of these findings in other settings. Regardless, the fact that we observed such variation even in a relatively advanced hospital setting indicates that there is much work to be done in improving the acceptability of AI to clinicians.

\section{Conclusion}

To our knowledge, this paper is one of the first to rigorously assess clinicians' interactions with a real AI system that predicts the effects of treatment strategies under uncertainty. This form of AI aims to complement human decision-makers by revealing previously-unseen patterns in historical outcomes, in contrast to deep learning models that are simply designed to save clinician effort by recapitulating human decision-making. While many clinicians in our study were generally receptive to the idea of AI support, the ones who found the AI Clinician most useful in practice were those who saw it as a source of additional evidence—a piece of data that could inform their decision alongside their assessment. Reshaping these AI tools as a source of individually-validated recommendations may be one way to clarify their intended use and to facilitate evaluation of their impacts on decisions in the process. Together with advances in human-centered algorithm design and more nuanced decision metrics, we envision this work as a step towards AI-driven prediction tools that foster a refined notion of ``appropriate reliance.''

\begin{acks}
We thank Ziyang Guo, Claire Chen, and Medha Palavalli for contributions to the modeling and visualization code; Billie Davis for assistance with transcription; and Dr. Emily Brant, Alex Cabrera, Nur Yildirim, Dominik Moritz, and John Zimmerman for helpful discussions around the manuscript. We also thank the numerous clinicians who participated in pilots and study sessions. This work was supported by a research grant from the United States National Institutes of Health (R35HL144804), by a National Science Foundation Graduate Research Fellowship (DGE2140739), and by the Carnegie Mellon University Center of Machine Learning and Health.
\end{acks}

\bibliographystyle{ACM-Reference-Format}
\bibliography{references}

% \appendix

% \section{Appendix Stuff}

\end{document}